\documentclass[12pt]{article}

\usepackage[utf8]{inputenc}
\usepackage[T1]{fontenc}
\usepackage[version=3]{mhchem}
\usepackage{graphicx}
\usepackage[justification=justified]{caption}
\usepackage{float}
\usepackage{amsmath, amssymb}
\usepackage{textcomp}
\usepackage{xcolor}
\usepackage[square,numbers,sort]{natbib}
\usepackage{authblk}
\usepackage[margin=2.5cm]{geometry}
\usepackage[switch]{lineno} 

\title{\LARGE \bfseries Bulk spin-orbit torque-driven spin Hall nano-oscillators using PtBi alloys}

\author[1]{Utkarsh Shashank}
\author[1,2,3]{Akash Kumar}
\author[4,5,6]{Tahereh Sadat Parvini}
\author[4]{Hauke Heyen}
\author[7]{Lunjie Zeng}
\author[7]{Andrew B. Yankovich}
\author[8]{Mona Rajabali}
\author[7]{Eva Olsson}
\author[4]{Markus Münzenberg\thanks{markus.muenzenberg@uni-greifswald.de}}
\author[1,2,3]{Johan Åkerman\thanks{johan.akerman@physics.gu.se}}

\affil[1]{Applied Spintronics Group, Department of Physics, University of Gothenburg, 412 96 Gothenburg, Sweden}
\affil[2]{Research Institute of Electrical Communication, Tohoku University, Sendai, Japan}
\affil[3]{Center for Science and Innovation in Spintronics, Tohoku University, Sendai, Japan}
\affil[4]{Institut für Physik, Universität Greifswald, 17489 Greifswald, Germany}
\affil[5]{Walther-Meißner-Institut, Bayerische Akademie der Wissenschaften, 85748 Garching, Germany}
\affil[6]{Munich Center for Quantum Science and Technology (MCQST), Munich, Germany}
\affil[7]{Department of Physics, Chalmers University of Technology, 412 96 Gothenburg, Sweden}
\affil[8]{NanOsc AB, Kista, Sweden}

\date{}

\DeclareUnicodeCharacter{2009}{\,}

\begin{document}

\maketitle

\textbf{Keywords:} spin-orbit torque, spin Hall effect, spin Hall nano-oscillator, auto-oscillation, extrinsic side-jump scattering

\begin{abstract}
Spin-orbit-torque-driven auto-oscillations in spin Hall nano-oscillators (SHNOs) offer a transformative pathway toward energy-efficient, nanoscale microwave devices for next-generation neuromorphic computing and high-frequency technologies. A key requirement for achieving robust, sustained oscillations is reducing the threshold current ($I_{\text{th}}$), strongly governed by spin Hall efficiency ($\theta_{\text{SH}}$). However, conventional strategies to enhance $\theta_{\text{SH}}$ face trade-offs, including high longitudinal resistivity, interfacial effects, and symmetry-breaking torques that limit performance. Here, we demonstrate a substantial enhancement of the bulk spin Hall effect in PtBi alloys, achieving over a threefold increase in $\theta_{\text{SH}}$, from 0.07 in pure Pt to 0.24 in Pt$_{94.0}$Bi$_{6.0}$ and 0.19 in Pt$_{91.3}$Bi$_{8.7}$, as extracted from DC-bias spin-torque ferromagnetic resonance. The enhanced $\theta_{\text{SH}}$ originates from bulk-dominated, extrinsic side-jump scattering across all PtBi compositions. Correspondingly, we observe a 42\% and 32\% reduction in $I_{\text{th}}$ in 100 nm SHNOs based on Co$_{40}$Fe$_{40}$B$_{20}$(3 nm)/Pt$_{94.0}$Bi$_{6.0}$(4 nm) and Co$_{40}$Fe$_{40}$B$_{20}$(3 nm)/Pt$_{91.3}$Bi$_{8.7}$(4 nm), respectively. Structural characterization reveals reduced Pt crystallinity, along with emergence of preferred crystallographic orientations upon introducing higher Bi concentrations. Together, these results position PtBi alloys as a compelling alternative to conventional 5$d$ transition metals, enabling enhanced $\theta_{\text{SH}}$ and significantly lower $I_{\text{th}}$, thus opening new avenues for energy-efficient neuromorphic computing and magnetic random access memory.
\end{abstract}

\section{Introduction}
The efficient control of magnetization via spin-orbit torque (SOT) \cite{shao2021roadmap}, primarily generated by the Spin Hall effect (SHE)~\cite{hirsch1999spin,dyakonov1971current}, lies at the core of modern spintronics~\cite{dieny2020natelectron}. It plays a key role in magnetization switching~\cite{liu2012science, yu2014switching, miron2011perpendicular, cai2022first}, in the generation and control of propagating spin waves~\cite{fulara2019spin,kumar2025spin}, in spin Hall nano-oscillators (SHNOs) \cite{demidov2014nanoconstriction,kumar2024mutual}, and in the movement of  domain walls and skyrmions \cite{emori2013current, bernstein2025spin, khvalkovskiy2013matching}. In a heavy metal (HM) with large spin-orbit coupling, the SHE generates a sizeable transverse spin current density, $\mathbf{j}_s$, from a longitudinal charge current density, $\mathbf{j}_c$. The generated $\mathbf{j}_s$ can 
exert different torques on the magnetization of an adjacent ferromagnet (FM) layer, where the damping-like torque, $\boldsymbol{\tau}_{\text{DL}}$, 
is the component collinear with the usual damping, $\boldsymbol{\tau}_{\text{$\alpha$}}$, of the FM layer~\cite{liu2011spin}. 
If $\boldsymbol{\tau}_{\text{DL}}$ balances $\boldsymbol{\tau}_{\text{$\alpha$}}$, the threshold condition for FM auto-oscillations (AO), with a corresponding threshold current ($I_{\text{th}}$), is reached in the SHNO~\cite{liu2013spectral,kasai2014modulation}. 

Over the past decade, SOT-driven nanoconstriction-based SHNOs have garnered significant attention thanks to their 
straightforward nano-fabrication, direct optical and gate access to the auto-oscillating FM region, and a rich magneto-dynamical behavior~\cite{chen2016spin}. SHNOs have shown tremendous promise for mutual synchronization in chains \cite{awad2017long,kumar2023robust} and arrays \cite{zahedinejad2020two,behera2024ultra,behera2025ultra}, paving the way for applications in neuromorphic computing \cite{zahedinejad2022memristive,torrejon2017neuromorphic,grollier2020neuromorphic}, Ising machines \cite{houshang2022phase}, magnonic conduits~\cite{kumar2025spin} and high frequency GHz technologies \cite{choi2022voltage}. 

To reduce $I_{\text{th}}$, a higher conversion of $\mathbf{j}_c$ to $\mathbf{j}_s$, defined as the spin Hall efficiency, $\theta_{\text{SH}}$, is required. $\theta_{\mathrm{SH}}$ can be expressed as $\mathbf{j}_{\mathit{s}} = \frac{\hbar}{2e} \, \theta_{\mathrm{SH}} \left( \mathbf{j}_{\mathit{c}} \times \hat{\boldsymbol{\sigma}} \right)$, where $\hat{\boldsymbol{\sigma}}$ is the polarization of the spin current, $e$ is the elementary charge, and $\hbar$ is the reduced Planck constant, and the most commonly used HM Pt has a moderate $\theta_{\text{SH}}$ $\approx$ 0.05-0.09, from its intrinsic SHE \cite{wang2014determination}. Ion-implantation \cite{shashank2021enhanced,shashank2021highly,shashank2023disentanglement,shashank2025giant} or sputtering \cite{yang2021maximizing,demasius2016enhanced,chen2017tunable} of non-metallic lighter impurities, such as S, O, N, P can also be used to tune the $\theta_{\text{SH}}$ of Pt, Ta, W. However, such tuning requires careful materials engineering and often comes with a trade-off in the form of increased longitudinal resistivity (\(\rho_{\mathrm{xx}}\)), as observed in TaN, which can exhibit values as high as \(3000~\mu\Omega\cdot\text{cm}\)~\cite{chen2017tunable}. In contrast, metallic impurities introduced by co-sputtering with the HM, neither require complex engineering nor introduce drastic changes in $\rho_{\text{xx}}$. A better trade-off between $\theta_{\text{SH}}$ and $\rho_{\text{xx}}$, is crucial for low-power consumption in SOT magnetic random access memory (SOT-MRAM) \cite{shashank2025giant,wang2022giant}. In this context, Pt, Ta, and W based alloys have been explored ($\theta_{\text{SH}}$ values in parentheses), such as Pt$_{28}$Cu$_{72}$ (0.07) \cite{ramaswamy2017extrinsic}, Pt$_{45}$Pd$_{55}$ (0.06) \cite{zhou2016disentanglement}, Pt$_{92}$Bi$_{8}$ (0.10) \cite{hong2018giant}; W$_{100-x}$Ta$_{x}$ (-0.35 to -0.62) \cite{behera2022energy}, and Cu$_{100-x}$Ta$_{x}$ (-0.04 to -0.09) \cite{chen2017tunable}. Within this framework, Hayashi \textit{et al.} showed that Pt$_{100-x}$Bi$_x$/Co bilayers with substantial Bi concentrations (x = 25–50)  exhibit markedly improved $\theta_{\text{SH}}$ and reduced switching current densities relative to pure Pt, identifying PtBi as a strong candidate for energy-efficient magnetization switching \cite{chi2021charge}. Complementing this, Münzenberg \textit{et al.}~demonstrated that Pt$_{92}$Bi$_8$-based hetero-structures can act as high-performance THz spintronic emitters, delivering broader bandwidths and higher central frequencies than conventional bilayer systems \cite{winkel2024comparative}. In addition, 
at higher Bi concentrations—such as in stoichiometric PtBi$_2$—this material has attracted attention as a topological system, exhibiting giant three-dimensional Rashba-like spin splitting ($\alpha_\mathrm{R} \approx 4.36~\text{eV}\cdot\text{\AA}$) \cite{feng2019rashba} and surface-bound superconducting gaps reaching up to 20 meV \cite{schimmel2024surface}. Despite these advances, a unified understanding of the reciprocal connection between $\theta_{\text{SH}}$ and $I_{\text{th}}$ in Pt$_{100-x}$Bi$_x$ alloys—via both ST-FMR and direct AO measurements—remains unexplored.

Here, we report on a dramatic increase in $\theta_{\text{SH}}$ by more than three times when alloying Pt with Bi, and a corresponding decrease of $I_{\text{th}}$ by $42\%$ in SHNOs based on Pt$_{94.0}$Bi$_{6.0}$ and Co$_{40}$Fe$_{40}$B$_{20}$. We deposited Pt$_{100-x}$Bi$_{x}$ alloy thin films using electron-beam co-evaporation and systematically varied the Bi concentrations across $x$= 0.0, 3.9, 6.0, and 8.7 (hereafter denoted as Pure Pt, Pt$_{96.1}$Bi$_{3.9}$, Pt$_{94.0}$Bi$_{6.0}$, and Pt$_{91.3}$Bi$_{8.7}$, respectively). Structural characterization by grazing-incidence X-ray diffraction (GIXRD) reveals a progressive loss of crystallinity with increasing Bi content, suggesting enhanced scattering contributions. Cross-sectional transmission electron microscopy (TEM) indicates a reduction in crystallinity of Pt, and a uniform distribution of Bi in Pt. Linewidth analysis of DC-biased spin-torque ferromagnetic resonance (ST-FMR) measurements \emph{vs.}~current yields a sharp increase in $\theta_{\text{SH}}$ from $0.07 \pm 0.01$ in pure Pt to $0.24 \pm 0.02$ and $0.19 \pm 0.01$ in the Pt$_{94.0}$Bi$_{6.0}$ and Pt$_{91.3}$Bi$_{8.7}$ alloys, respectively. To rule out artifacts and confirm the bulk origin of SOT, we demonstrate a clear $\sin2\phi\cos\phi$ in-plane angular dependence of the SOT for all PtBi alloys. Thanks to the increase in $\theta_{\text{SH}}$, we observe a $42\%$ and $32\%$ reduction in $I_{\text{th}}$ for Pt$_{94.0}$Bi$_{6.0}$ and Pt$_{91.3}$Bi$_{8.7}$ based 100 nm SHNOs. Our findings of a higher $\theta_{\text{SH}}$ driven by bulk-extrinsic SHE, accompanied by a reduced $I_{\text{th}}$, establish Bi alloying in Pt as a robust strategy for engineering energy-efficient SHNOs, opening new avenues beyond conventional $5d$ transition metals.

\section{Results and Discussion}

\subsection{Structural Characterization via GIXRD and TEM} 
\begin{figure}[H]
    \centering
    \includegraphics[width=0.9\linewidth]{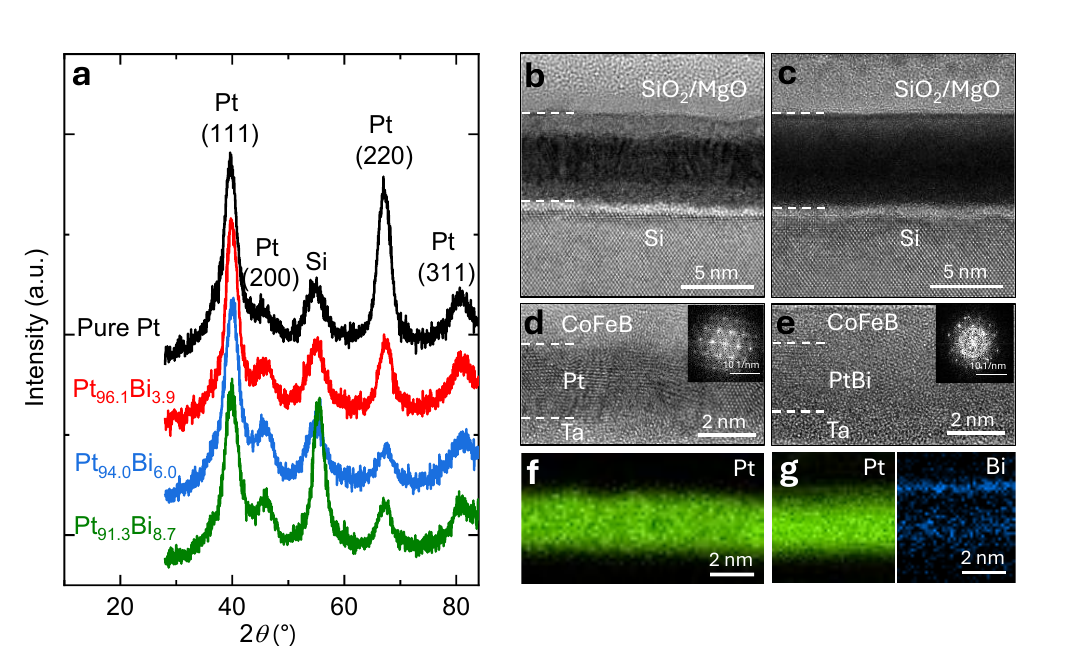}  
    \caption{(a) GIXRD patterns of the pure Pt stack (black) and PtBi alloy stacks with compositions Pt$_{96.1}$Bi$_{3.9}$ (red), Pt$_{94.0}$Bi$_{6.0}$ (blue), and Pt$_{91.3}$Bi$_{8.7}$ (green).  
    (b,c) Low-resolution cross-sectional bright-field TEM images of (b) the pure Pt stack (HR-Si/Ta(2.4~nm)/Pt(4.0~nm)/Co$_{40}$Fe$_{40}$B$_{20}$(3.0~nm) and (c) the Pt$_{94.0}$Bi$_{6.0}$ stack (HR-Si/Ta(2.4~nm)/Pt$_{94.0}$Bi$_{6.0}$(4.0~nm)/Co$_{40}$Fe$_{40}$B$_{20}$(3.0~nm)) both showing uniform, well-defined layers with comparable total thicknesses.  Both layers are capped with MgO(2.0~nm)/SiO$_2$(3.0~nm). 
    (d,e) High-resolution TEM images of the same stacks. (d) The pure Pt stack shows clear lattice fringes and sharp Bragg spots (inset FFT), indicative of a polycrystalline Pt structure. (e) The Pt$_{94.0}$Bi$_{6.0}$ stack displays reduced fringe contrast and weaker Bragg spots (inset), suggesting lower crystallinity or a change in preferred orientation.  
    (f,g) EDS elemental maps of Pt and Bi from HAADF-STEM for the pure Pt and Pt$_{94.0}$Bi$_{6.0}$ stacks, respectively. Bi is uniformly distributed within the Pt$_{94.0}$Bi$_{6.0}$ layer, with slight enrichment at the PtBi/Co$_{40}$Fe$_{40}$B$_{20}$ interface. Additional HAADF-STEM images and full elemental maps are provided in Supporting Information S1 and S2.}
    \label{fig:Figure/Figure0_TEM}
\end{figure}

\textbf{Figure~\ref{fig:Figure/Figure0_TEM}}a shows Grazing incidence X-ray Diffraction (GIXRD) patterns of Pt$_{100-x}$Bi$_x$ thin films (see Experimental Section). Reflections from the (111), (200), (220), and (311) planes confirm a polycrystalline face-centered cubic (\textit{fcc}) phase in pure Pt and Pt$_{96.1}$Bi$_{3.9}$~\cite{shashank2025giant,lau2021highly}. Increasing the Bi content reduces the overall reflection intensities, particularly that of the (220) peak, indicating a reduction in crystallinity \cite{shashank2025giant,shashank2024charge,hecq1981sputtering}. Peak broadening becomes more pronounced in Pt$_{94.0}$Bi$_{6.0}$ and Pt$_{91.3}$Bi$_{8.7}$. The intensity ratio \( I_{(111)}/I_{(220)} \) rises from 0.39 (Pure Pt) to 1.65 (6\% Bi), then falls to 1.07 at 8.7\% Bi. Similarly, the lattice parameter increases from 3.953 to 3.966~\AA{} (0–6\% Bi), followed by a slight decrease to 3.959~\AA{} at 8.7\% Bi. This non-monotonic trend suggests a shift in preferred crystallographic orientation, likely due to strain-induced texturing at higher Bi concentrations, consistent with previous report~\cite{chi2021charge}. 

 Figure~\ref{fig:Figure/Figure0_TEM}b,c display low-resolution cross-sectional bright-field TEM images of the pure Pt stack and the Pt$_{94.0}$Bi$_{6.0}$ stack, respectively. Both stacks exhibit well-defined, uniform layers tructures with comparable total thicknesses. Figures~\ref{fig:Figure/Figure0_TEM}d,e present high-resolution bright-field TEM images of the same samples. In the pure Pt sample (Figure~1d), clear lattice fringes are visible, and the fast Fourier transform (FFT) analysis (inset) shows distinct Bragg spots and a clear ring, indicating a polycrystalline structure of the Pt layer~\cite{shashank2021enhanced,shashank2025giant,peng2019tunable}. In contrast, the Pt$_{94.0}$Bi$_{6.0}$ sample (Figure~\ref{fig:Figure/Figure0_TEM}e) exhibits weaker lattice fringes and diminished Bragg spots (inset), suggesting reduced crystallinity or a shift in preferred crystallographic orientation, consistent with the GIXRD results. Figure~\ref{fig:Figure/Figure0_TEM}f,g provide Energy-dispersive X-ray spectroscopy (EDS) elemental maps from High-angle annular dark field scanning transmission electron microscopy (HAADF-STEM) for the same samples. In the Pt$_{94.0}$Bi$_{6.0}$ stack (Figure~\ref{fig:Figure/Figure0_TEM}g, right panel), Bi is uniformly distributed throughout the target Pt layer, with no sign of clustering. A modest enrichment of Bi is also observed at the Pt$_{94.0}$Bi$_{6.0}$/Co$_{40}$Fe$_{40}$B$_{20}$ interface. The impact of both bulk and interfacial Bi distributions on the SHE in Pt will be addressed in subsequent sections.

\subsection{Spin-torque ferromagnetic resonance measurements} 

\begin{figure}[H]
    \centering
    \includegraphics[width=1\linewidth]{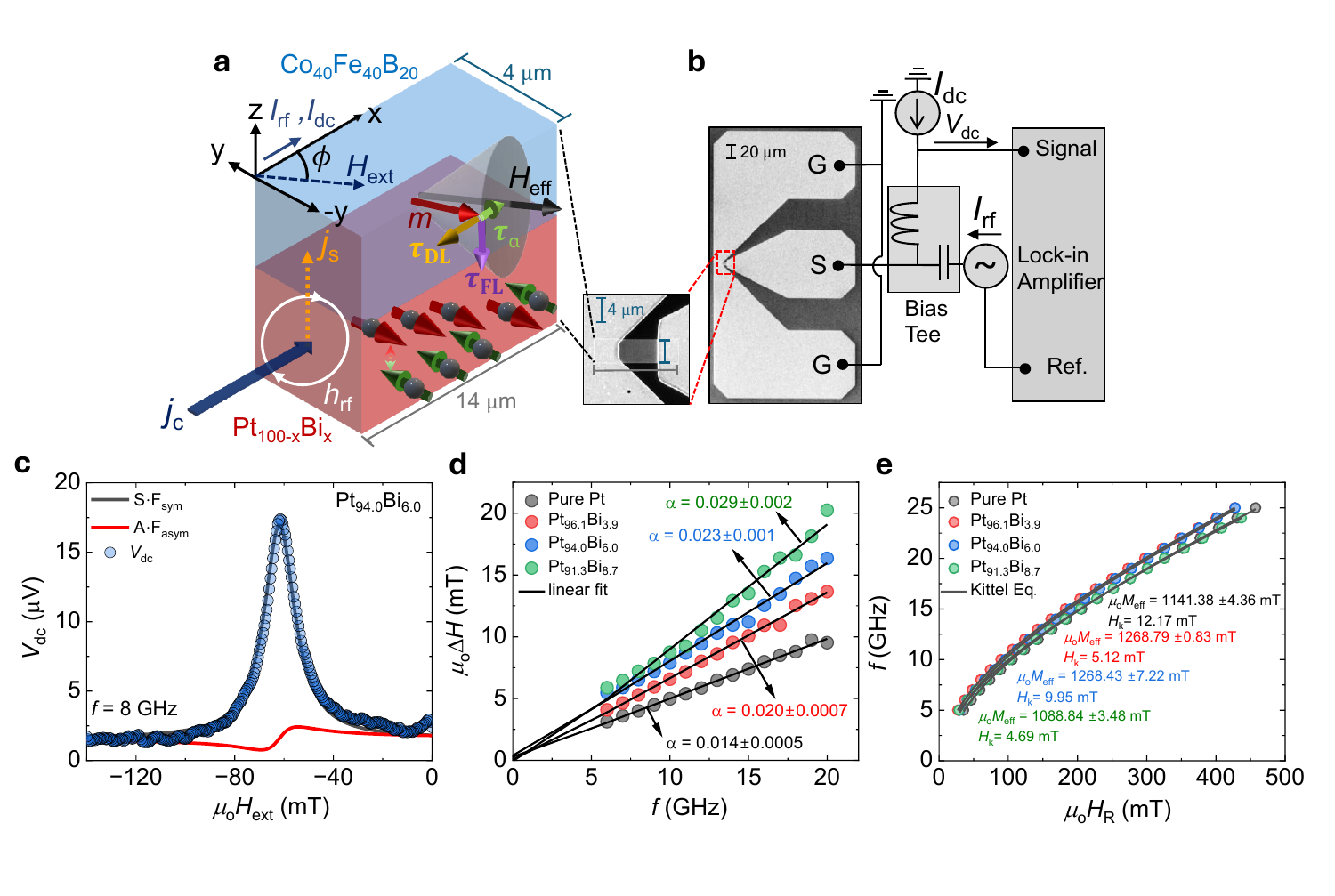} 
    \caption{a) Schematic of the SHE and the microscopic mechanism of the ST-FMR excitation, depicting the torques acting on the FM magnetization. b) The ST-FMR measurement technique and detection principle along with the micro-bar device connected to a GSG-CPW. A zoom-in SEM image of the red square 
    shows the micro-bar. c) ST-FMR voltage ($V_{\text{dc}}$) at $f$ = 8 GHz of Pt$_{94.0}$Bi$_{6.0}$, with separate symmetric and antisymmetric components fit to Eq.~(1). d) $\Delta H$ vs $f$ for $f$ = 6--20 GHz, with solid black lines being linear fits to Equation~(2). e) $f$ vs.~$H_{\text{R}}$ for all samples evaluated in this study. Solid lines in (e) represent fits to the Kittel Equation~(3). }
    \label{fig:Figure1_ST-FMR}
\end{figure}

To evaluate how Bi doping modulates the SHE in Pt, we performed spin-torque ferromagnetic resonance (ST-FMR) measurements \cite{liu2011spin} (see Experimental section for details). \textbf{Figure~\ref{fig:Figure1_ST-FMR}}a illustrates the schematic of the ST-FMR measurement and Figure~\ref{fig:Figure1_ST-FMR}b shows the experimental setup. An in-plane microwave current \( I_{\text{rf}} \) with a power of 4 dBm was applied along the x-axis of the HM/FM bilayer via a ground-source-ground (GSG) coplanar waveguide (CPW). The external magnetic field \( H_{\text{ext}} \) was swept \(\pm 500\,\mathrm{mT}\) at an angle \(\phi = 70^\circ\) between \( I_{\text{rf}} \) and \( H_{\text{ext}} \). The charge current density \( \mathbf{j}_c \) (x-axis) driven by \( I_{\text{rf}} \) generates a spin current density \( \mathbf{j}_s \) along the z-axis with spin polarization \( \hat{\boldsymbol{\sigma}} \) along the y-axis. This \( \mathbf{j}_s \) exerts an in-plane $\boldsymbol{\tau}_{\text{DL}}$ on the magnetization \( \mathbf{m} \) of the FM layer. Simultaneously, the Oersted field, $h_{\text{rf}}$, generated by \( I_{\text{rf}} \) acts as an out-of-plane Oersted field torque, $\boldsymbol{\tau}_{\text{OF}}$. Interfacial effects, rather than bulk mechanisms, can also induce out-of-plane field-like torques (\(\boldsymbol{\tau}_{\text{FL}}\)) arising from the Rashba effect, particularly in thinner FM layers ($\lesssim 1~\text{nm}$) interfaced with HM \cite{fan2014quantifying,nan2015comparison}, which may overlap with \(\boldsymbol{\tau}_{\text{OF}}\). When $\boldsymbol{\tau}_{\text{FL}}$ is small, as in bulk SHE heterostructures, the combined torques ($\boldsymbol{\tau}_{\text{DL}}$ and $\boldsymbol{\tau}_{\text{OF}}$) govern the magnetization precession. The precession of \( \mathbf{m} \) around the effective field \( H_{\text{eff}} \) induces a time-dependent resistance variation via the anisotropic magnetoresistance (AMR), \( \Delta R \propto \cos^2\phi \) (see Supporting Information S3). Mixing with \( I_{\text{rf}} \), the varying $\Delta R$ generates a rectified ST-FMR voltage, \( V_{\text{dc}} \). The $I_{\text{rf}}$ was amplitude modulated at 98.76 Hz, serving as a reference signal for lock-in detection (see Figure~\ref{fig:Figure1_ST-FMR}b).

\setlength{\parindent}{15pt} Figure~\ref{fig:Figure1_ST-FMR}c shows the ST-FMR voltage spectrum measured at \textit{f}= 8 GHz and $I_{\text{dc}}$= 0 mA for Pt$_{94.0}$Bi$_{6.0}$. The measured $V_{\text{dc}}$ is defined as~\cite{liu2011spin}: \[V_{\text{dc}} = S\!F_{\text{sym}}(H_{\text{ext}}) + A\!F_{\text{asym}}(H_{\text{ext}}), \tag{1}\] where, $F_{\text{sym}}(H_{\text{ext}}) = \frac{(\Delta H)^2}{(\mu_0H_{\text{ext}} - \mu_0H_{\text{R}})^2 + (\Delta H)^2}$ is the symmetric (Lorentzian) component, and $F_{\text{asym}}(H_{\text{ext}}) = \frac{\Delta H (\mu_0H_{\text{ext}} - \mu_0H_{\text{R}})}{(\mu_0H_{\text{ext}} - \mu_0H_{\text{R}})^2 + (\Delta H)^2}$ is the antisymmetric (dispersive) component. The parameters $S$ and $A$ are the weight factors corresponding to the symmetric and antisymmetric components, respectively. $\Delta H$ and $H_{\text{R}}$ are the linewidth (half-width-at-half-maximum) and resonance field of the ST-FMR spectra, respectively. Strikingly, a very high symmetric component $S\!F_{\text{sym}}(H_{\text{ext}})$ confirming the high $\boldsymbol{\tau}_{\text{DL}}$ is observed for Pt$_{94.0}$Bi$_{6.0}$. 
Furthermore, in-plane angular ST-FMR measurements reveal that both the $S$ and $A$ weight factors exhibit a clear $\sin2\phi\cos\phi$ dependence across all Pt$_{100-x}$Bi$_{x}$ compositions (see Supporting Information S4 for details). This angular dependence is a characteristic of a conventional bulk-SOT origin, free from torque symmetry-breaking, and rules out significant contributions from experimental artifacts~\cite{shashank2025giant,harder2016electrical,mohan2025observation}.

\setlength{\parindent}{15pt} Next, we extract the Gilbert damping parameter $\alpha$ by plotting $\Delta H$ as a function of $f$ (see Figure~\ref{fig:Figure1_ST-FMR}d), using the relation \[\Delta H = \Delta H_{\text{0}} + \frac{2 \pi \alpha}{\gamma \mu_0} f, \tag{2}\] where $\Delta H_{\text{0}}$ denotes the $f$-independent inhomogeneous linewidth broadening \cite{bainsla2022ultrathin}, and $\gamma$ is the gyromagnetic ratio. A systematic increase in $\alpha$ is observed from $0.014 \pm 0.0005$ (pure Pt) to $0.020 \pm 0.0007$, $0.023 \pm 0.001$ and $0.029 \pm 0.002$ for Pt$_{96.1}$Bi$_{3.9}$, Pt$_{94.0}$Bi$_{6.0}$, and Pt$_{91.3}$Bi$_{8.7}$, respectively. This trend reflects enhanced spin current generation, potentially arising from spin pumping \cite{hong2018giant} and/or an increase in SHE \cite{shashank2021highly,shashank2023disentanglement}. The extracted $\Delta H_{\text{0}}$ remains low and positive (0.2-2.1 mT), indicating excellent film quality \cite{shashank2021enhanced, shashank2021highly}. Figure~\ref{fig:Figure1_ST-FMR}e depicts the frequency dependence of the resonance field, $H_{\text{R}}$ using the Kittel formula: \[f = \frac{\gamma \mu_0}{2\pi} \sqrt{(H_{\mathrm{R}} + H_{\mathrm{K}})\left(H_{\mathrm{R}} + H_{\mathrm{K}} + M_{\mathrm{eff}}\right)}, \tag{3}\] where, $H_{\text{k}}$ is in-plane magnetic anisotropy field and $M_{\text{eff}}$ is effective demagnetization \cite{kumar2018large}. The extracted $M_{\mathrm{eff}}$ values—1141.4~mT (pure Pt), 1268.8~mT (Pt$_{96.1}$Bi$_{3.9}$), 1268.4~mT (Pt$_{94.0}$Bi$_{6.0}$), and 1088.8~mT (Pt$_{91.3}$Bi$_{8.7}$)—along with $H_{\mathrm{K}}$ values of 4–12~mT, confirm no significant variation in magnetic properties upon Bi incorporation.

\begin{figure}[H]
    \centering
    \includegraphics[width=1\linewidth]{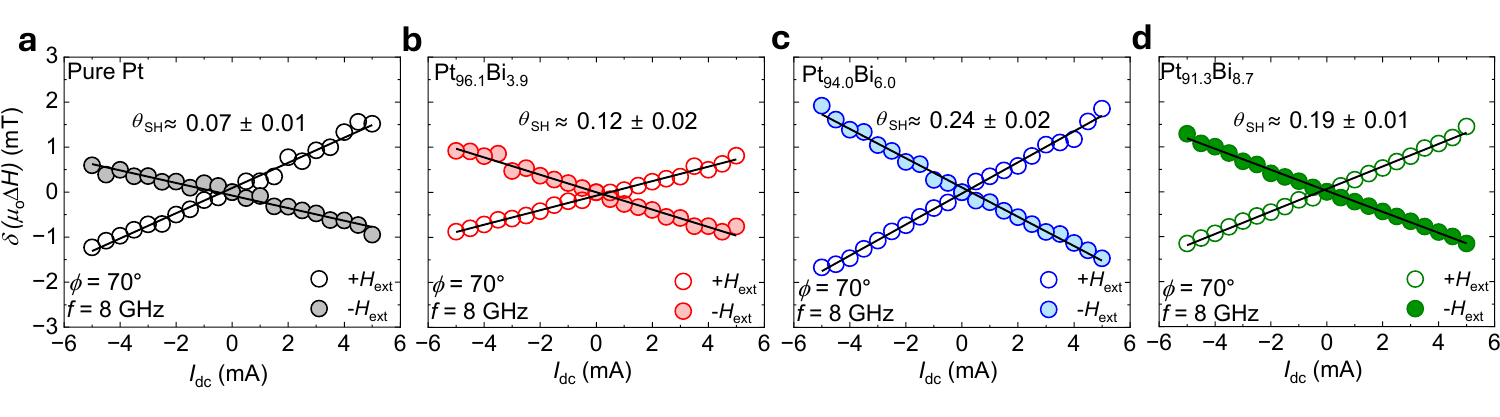}  
    \caption{Current-induced modulation of linewidth, $\delta(\mu_{\mathrm{o}} \Delta H)$ as a function of $I_{\text{dc}}$ at $f$ = 8 GHz for a) Pure Pt
    , b) Pt$_{96.1}$Bi$_{3.9}$, c) Pt$_{94.0}$Bi$_{6.0}$, and d) Pt$_{91.3}$Bi$_{8.7}$. Open circles correspond to positive magnetic field sweep ($+H_{\text{ext}}$), while filled circles correspond to negative sweep ($-H_{\text{ext}}$). The solid black lines 
are linear fits. The increasing slope of  $\delta(\mu_0\Delta H)$/$I_{\text{dc}}$ with Bi impurity in Pt, indicates enhanced $\theta_{\text{SH}}$.}
    \label{fig:Figure2_DC-bias ST-FMR}
\end{figure}

\setlength{\parindent}{15pt} To quantify $\theta_{\text{SH}}$, we use DC-bias ST-FMR \cite{liu2011spin}, where a direct current \(I_{\mathrm{dc}}\) is applied simultaneously with 
\(I_{\mathrm{rf}}\) (see Figure~\ref{fig:Figure1_ST-FMR}b) under an external magnetic field \(H_{\mathrm{ext}}\), to modulate the linewidth \(\Delta H\). \textbf{Figure~\ref{fig:Figure2_DC-bias ST-FMR}}a--d show the change in linewidth, $\delta(\mu_0\Delta H)$, as a function of \(I_{\mathrm{dc}}\) at \(f = 8\,\mathrm{GHz}\), measured at an angle \(\phi = 70^\circ\) with an input power of 4 dBm. Here, \(\phi\) is the angle between the combined current direction \((I_{\mathrm{rf}} + I_{\mathrm{dc}})\) and the external field \(H_{\mathrm{ext}}\). The angle \(\phi = 70^\circ\) was chosen to maximize the linewidth modulation since \(\sin \phi\) reaches near its peak between \(70-75^\circ\), and the AMR also exhibits significant variation within the \(45^\circ\text{--}75^\circ\) range~\cite{kasai2014modulation,wang2018fmr} fruitful for a clean ST-FMR spectrum (see Equation~4). The change in linewidth is defined as $\delta(\mu_0\Delta H)$= $\mu_0\Delta H \big|_{I_{\mathrm{dc}}}$-$\mu_0\Delta H \big|_{I_{\mathrm{dc}}=0}$, which varies linearly with $I_{\text{dc}}$. We then define the effective current-dependent damping parameter, $\alpha_{\mathrm{eff}}$ given as $\alpha_{\mathrm{eff}}$= $\frac{\gamma}{2\pi f} \delta(\mu_0\Delta H)$. Reversing the polarity of $H_{\text{ext}}$ from (+$H_{\text{ext}}$) $\phi$=$70^\circ$ to (-$H_{\text{ext}}$) at $\phi$=$250^\circ$ reverses the magnetization direction of Co$_{40}$Fe$_{40}$B$_{20}$, which correspondingly reverses the slope of $\delta(\mu_0\Delta H)$/$I_{\text{dc}}$, confirming the conventional SHE induced SOT \cite{kondou2021giant}. Using a parallel resistor model (see Supporting Information S5 for details on calculating \(\rho_{\text{xx}}\)), we estimate the current density in the HM layer, \(j_{\mathrm{dc,HM}}\), and extract \(\theta_{\text{SH}}\) from the slope \(\frac{\Delta \alpha_{\mathrm{eff}}}{\Delta j_{\mathrm{dc,HM}}}\) according to \cite{liu2011spin}:
\[\theta_{\text{SH}} = \left[ \frac{2e}{\hbar} \left( H_0 + \frac{M_{\mathrm{eff}}}{2} \right) \mu_0 M_{\mathrm{s}} t \left| \frac{\Delta \alpha_{\mathrm{eff}}}{\Delta j_{\mathrm{dc,HM}}} \right| \right] / \sin\phi, \tag{4}\] where $M_{\mathrm{s}}$ and $t$ are the saturation magnetization and thickness of Co$_{40}$Fe$_{40}$B$_{20}$, respectively, and $M_{\text{eff}}$ is the effective demagnetization obtained from Kittel fitting \cite{bainsla2022ultrathin,Kittel1948} (see Figure 2e). We observe a slight increase in $\theta_{\text{SH}}$ from $0.07 \pm 0.01$ (Pure Pt) to $0.12 \pm 0.02$ (Pt$_{96.1}$Bi$_{3.9}$), with a small doping of 3.9\%. However, increasing the Bi concentration to 6.0 \% and 8.7 \% yields a remarkable enhancement in $\theta_{\text{SH}}$ to $0.24 \pm 0.02$ (Pt$_{94.0}$Bi$_{6.0}$) and $0.19 \pm 0.01$ (Pt$_{91.3}$Bi$_{8.7}$). In particular, the highest $\theta_{\text{SH}}$ occurs at higher Bi concentrations where Pt loses its crystalline texture, as seen in Figure 1. This supports the correlation between reduced crystallinity and enhanced $\theta_{\text{SH}}$ in \textit{5d} transition metals. Recent experimental studies by Parkin \textit{et al.} \cite{wang2022giant} and Fukuma \textit{et al.} \cite{shashank2025giant} have similarly demonstrated that $\theta_{\text{SH}}$ peaks near the crystalline-amorphous boundary, where extrinsic scattering from impurities is maximized, which we will explore in later sections.\\

\setlength{\parindent}{0pt}

\subsection{Spin Hall nano-oscillators} 
\begin{figure}[H]
    \centering
    \includegraphics[width=1\linewidth]{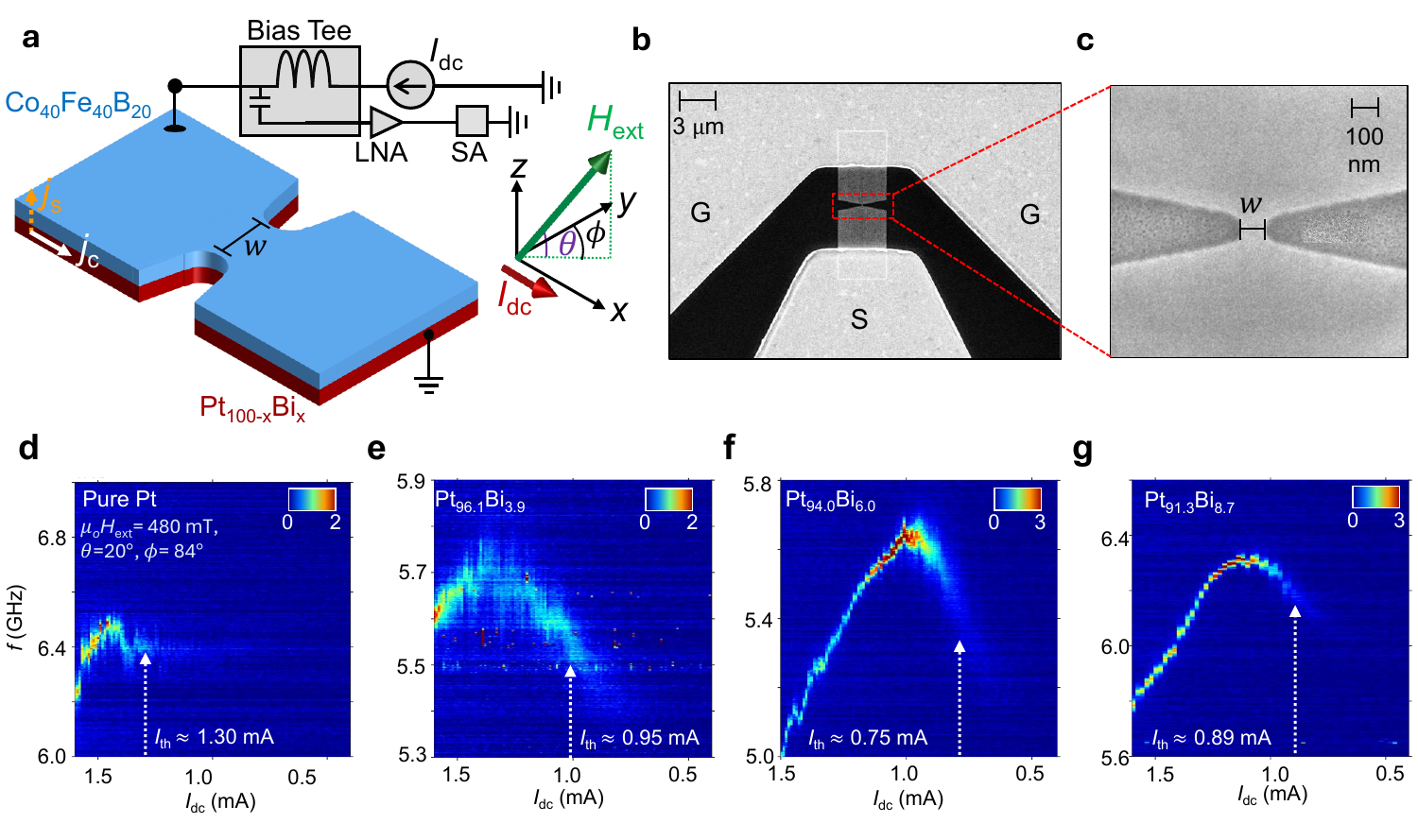}  
    \caption{a) Schematic of the SHNO and measurement setup. b) SEM image of the GSG-based CPW with SHNO. c) Enlarged SEM of the 100 nm wide SHNO (black scale bars in b,c). PSD from the 100 nm SHNO as a function of \(I_{\text{dc}}\) at \(\mu_0 H_{\text{ext}} = 480\) mT, \(\phi=20^\circ\), \(\theta=84^\circ\) for d) Pure Pt, e) Pt\(_{96.1}\)Bi\(_{3.9}\), f) Pt\(_{94.0}\)Bi\(_{6.0}\), and g) Pt\(_{91.3}\)Bi\(_{8.7}\). Color bars indicate peak power (dB above noise floor). The white arrows mark respective \(I_{\text{th}}\).}
    \label{fig:Figure3_SHNOschematic}
\end{figure}

We fabricated 100 nm wide SHNOs (see Experimental Section for details) and measured AO on all Pt$_{100-x}$Bi$_{x}$/Co$_{40}$Fe$_{40}$B$_{20}$ hetero-structures to observe the effect of a higher $\theta_{\text{SH}}$ on $I_{\text{th}}$. A schematic of the setup is shown in \textbf{Figure~\ref{fig:Figure3_SHNOschematic}}a. A direct current, $I_{\text{dc}}$ (along the x-axis) was applied through a Bias Tee, while the resulting microwave signals were extracted through the high-frequency port, amplified using a low-noise amplifier (LNA), and analyzed with a spectrum analyzer (SA). A moderate out-of-plane field ($\theta$$=$84$^\circ$) was selected to achieve a weak negative nonlinearity~\cite{dvornik2018origin}. The in-plane angle, $\phi$$=$20$^\circ$, was chosen to ensure adequate electrical sensitivity of the signal ($\phi$ is the angle between the y-axis and $H_{\text{ext}}$).

\setlength{\parindent}{15pt} Figure 4b shows the scanning electron microscope (SEM) image of the device under test, and Figure 4c presents the enlarged SEM image of the 100 nm wide SHNO. The actual measured width $w$ is in good agreement with the nominal design value. To understand the reciprocal relationship between the enhancement in $\theta_{\text{SH}}$ and the reduction in $I_{\text{th}}$, we performed current-dependent AO and extracted the Power Spectral Density (PSD) generated by the 100 nm wide SHNO under $\mu_o$$H_{\text{ext}}$ = 480 mT ($\phi$$=$20$^\circ$ and $\theta$$=$84$^\circ$) for all samples, as shown in Figure 4d-g. A non-monotonic current dependence of the AO frequency is observed, attributed to a localized AO ``bullet'' mode promoted by negative nonlinearity, $\mathcal{N}$, 
typical for a FM layer with its magnetization far from the film normal \cite{dvornik2018origin}. The $I_{\text{th}}$ is 1.30 mA for pure Pt, decreasing to 0.95 mA for Pt$_{96.1}$Bi$_{3.9}$. At higher Bi concentrations, $I_{\text{th}}$ drops significantly to 0.75 mA ($42\%$ reduction) for Pt$_{94.0}$Bi$_{6.0}$ and 0.89 mA ($32\%$ reduction) for Pt$_{91.3}$Bi$_{8.7}$. Concurrently, the peak and integrated output power increase while the linewidth decreases at higher Bi concentrations compared to pure Pt. See Supporting Information S6 for AMR, S7 for AO at other fields, S8 for $I_{\text{th}}$ extraction, and S9 for peak power, output power, and linewidth.\\

\setlength{\parindent}{0pt}

\subsection{Reciprocity and mechanism of spin Hall effect in PtBi alloys} 
\begin{figure}[H]
    \centering
    \includegraphics[width=1\linewidth]{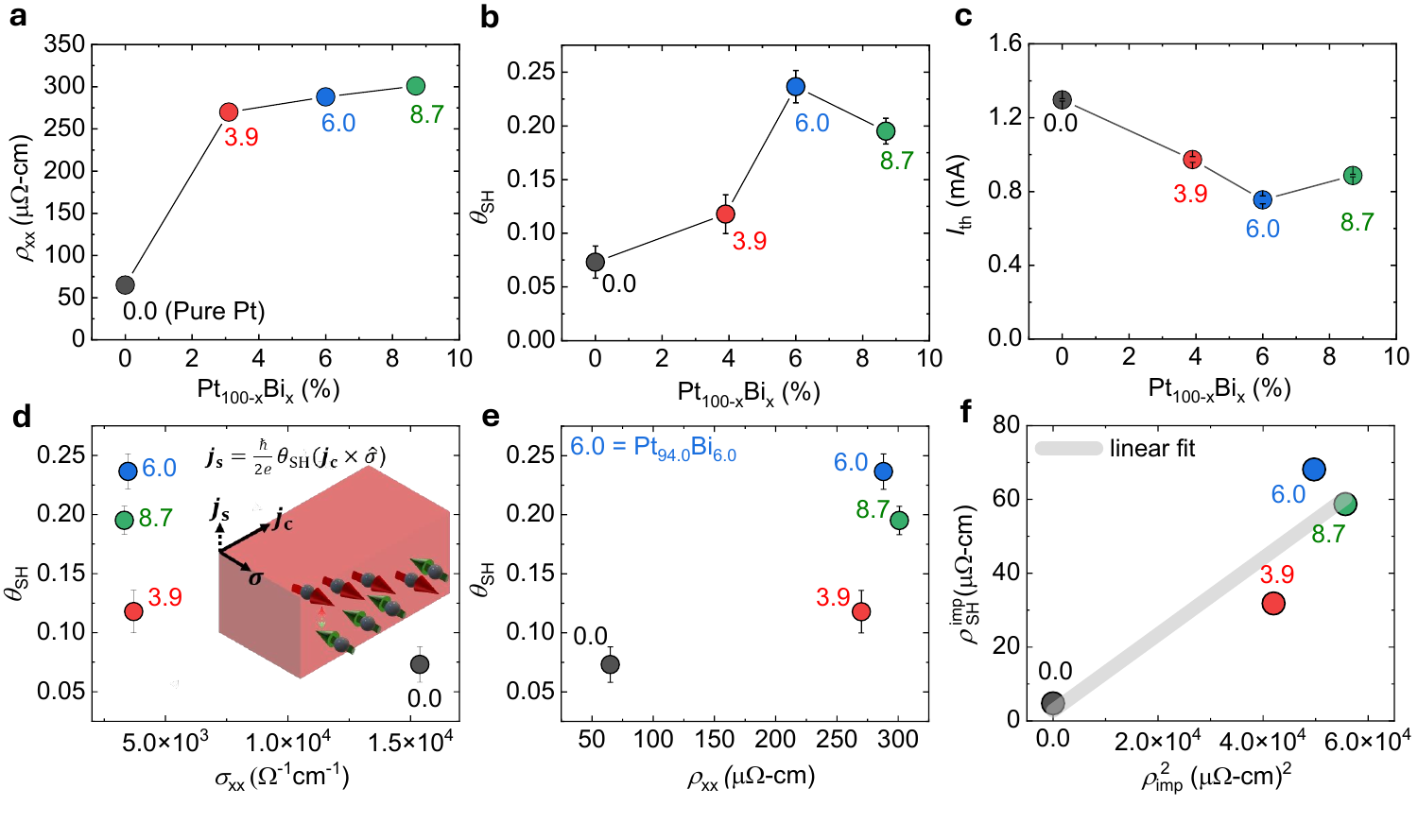}  
    \caption{a) \(\rho_{\text{xx}}\), b) \(\theta_{\text{SH}}\), and c) \(I_{\text{th}}\) versus Bi content \(x\) in Pt\(_{100-x}\)Bi\(_x\) (wt.\%). Error bars in (b) represent standard error from equation~2 and 4; in (c), mean deviation (MD) from \(I_{\text{th}}\) datasets (Supporting Information S7 for datasets; see Experimental Section for MD details). \(\theta_{\text{SH}}\) plotted against d) \(\sigma_{\mathrm{xx}}\) and e) \(\rho_{\text{xx}}\). Inset in (d) shows bulk SHE schematic in PtBi alloy. f) \(\rho_{\text{imp}}^{\text{SH}}\) as a function of \(\rho_{\text{imp}}^2\) with linear fits as a solid line.}
    \label{fig:Figure5_parameter}
\end{figure}

Building on the structural analysis and the observed reciprocity between ST-FMR and AO, several key parameters—\(\rho_{\text{xx}}\), \(\theta_{\text{SH}}\), and \(I_{\text{th}}\)— are analyzed across all Pt\(_{100-x}\)Bi\(_x\) alloys (see \textbf{Figure~\ref{fig:Figure5_parameter}}a-c). For pure Pt and 3.1\% Bi doping, \(\theta_{\text{SH}}\) increases from 0.07 to 0.12, with a corresponding rise in \(\rho_{\text{xx}}\) from 65 to 270 \(\mu\Omega\cdot\text{cm}\), reducing \(I_{\text{th}}\) from 1.30 mA to 0.97 mA. At 6.0\% Bi, \(\rho_{\text{xx}}\) rises marginally to 288 \(\mu\Omega\cdot\text{cm}\), while \(\theta_{\text{SH}}\) significantly rises to 0.24, yielding the lowest \(I_{\text{th}}\) of 0.75 mA. At 8.7\% Bi, \(\rho_{\text{xx}}\) further increases to 301 \(\mu\Omega\cdot\text{cm}\), but \(\theta_{\text{SH}}\) saturates and decreases to 0.19, causing \(I_{\text{th}}\) to slightly rise to 0.89 mA. This interplay between $\theta_{\mathrm{SH}}$ and $\rho_{\mathrm{xx}}$ is critical for minimizing power consumption, expressed by the factor, $\rho_{\mathrm{xx}} / \theta_{\mathrm{SH}}^2$ relevant for SOT-MRAM applications (Supporting Information S5) \cite{shashank2025giant,wang2022giant}. The output power is visibly higher at 6\% and 8.7\% Bi impurity in Pt compared to pure Pt (Figure~4d--g). Notably, linewidth narrows to $\approx$ \(25\ \text{MHz}\) at these higher Bi concentrations, likely due to increased \(\rho_{\text{xx}}\) redirecting current through Co\(_{40}\)Fe\(_{40}\)B\(_{20}\) enhancing output power and reducing linewidth (Supporting Information S9). 
Furthermore, despite the increase in $\alpha$ with Bi doping, the enhanced $\boldsymbol{\tau}_{\text{DL}}$ from the SHE compensates for $\boldsymbol{\tau}_{\alpha}$, leading to a reduction in $I_{\text{th}}$. This reduction would likely have been more significant if $\alpha$ had remained constant or decreased (see Figure~2d), complicating the direct assessment of SOT effects. The actual effect of torque on the AO is thus likely underestimated, and FM layer with lower $\alpha$, such as GdFeCo \cite{bainsla2022ultrathin} could further enhance the reduction in $I_{\text{th}}$.

\setlength{\parindent}{15pt} To elucidate the mechanism underlying the SHE, we examine the electrical conductivity \(\sigma_{\mathrm{xx}}\). Figure~5d shows \(\theta_{\text{SH}}\) \emph{vs.}~\(\sigma_{\mathrm{xx}}\), with values lying in the bad-metal/dirty-metal regime (\(\sigma_{\mathrm{xx}} \lesssim 10^{4} \ \Omega^{-1}\text{cm}^{-1}\)). In this 
regime, a linear dependence of \(\theta_{\text{SH}}\) on \(\sigma_{\mathrm{xx}}\) indicates a dominant intrinsic SHE contribution in Pt as demonstrated by Casanova \textit{et al.}~\cite{sagasta2016tuning}. However, no clear linear trend is observed, suggesting coexistence of intrinsic and extrinsic side-jump scattering mechanisms \cite{shashank2025giant,shashank2024charge,sagasta2016tuning}. Moreover, the increase of \(\theta_{\text{SH}}\) with \(\rho_{\text{xx}}\) (Figure~5e) supports a dominant role of side-jump scattering and/or disorder-enhanced intrinsic contributions \cite{chen2017tunable}. In metals with strong spin-orbit coupling, the anomalous Hall effect (AHE) scales with \(\rho_{\text{xx}}\) either linearly (extrinsic skew scattering) or quadratically (extrinsic side-jump scattering or intrinsic mechanisms) \cite{hao2015giant}. Recent studies have established that SHE shares this scaling behavior \cite{sinova2015revmodern}. The observed quadratic dependence of \(\theta_{\text{SH}}\) on \(\rho_{\text{xx}}\) hints at the contribution of bulk-extrinsic side-jump scattering  (Figure~5e) and rules out significant interfacial Bi effects on SHE, as confirmed by TEM and consistent with bulk alloy characteristics. To further confirm the mechanism, the intrinsic SHE contribution was subtracted, revealing \(\rho_{\text{imp}}^{\text{SH}} \propto \rho_{\text{imp}}^{2}\) (Figure~5f), indicative of extrinsic side-jump scattering dominance \cite{shashank2021highly,shashank2023disentanglement,ramaswamy2017extrinsic,shashank2024charge,kumar2018large,asomoza1983gadolinium}. Here, \(\rho_{\text{imp}}^{\text{SH}}\) denotes the impurity-induced spin Hall resistivity, and \(\rho_{\text{imp}}\) the resistivity due to impurities (See Supporting Information S10 for details on mechanism of SHE, including the exclusion of skew scattering as a contributing mechanism, and the extraction method for $\rho_{\text{imp}}^{\text{SH}}$ and $\rho_{\text{imp}}$). Therefore, increasing Bi concentration in Pt enhances extrinsic side-jump scattering, substantially boosting \(\theta_{\text{SH}}\) and reducing the threshold current \(I_{\text{th}}\) necessary to drive robust AO, thereby enabling energy-efficient SHNOs.

\section{Conclusion}
In summary, we demonstrate that Bi doping in Pt enhances $\theta_{\text{SH}}$ by over threefold and substantially reduces the threshold current $I_{\text{th}}$ in SHNOs, enabling a superior $\rho_{\text{xx}}$ trade-off crucial for low-power spintronic applications. The concurrent enhancement of $\theta_{\text{SH}}$ and reduction in $I_{\text{th}}$ underscores the reciprocal relationship between the spin Hall effect and auto-oscillations, as revealed by complementary ST-FMR and AO measurements. Furthermore, angular-dependent ST-FMR confirms a $\sin2\phi\cos\phi$ symmetry across all Pt$_{100-x}$Bi$_{x}$ alloys, indicative of a bulk-origin, conventional spin–orbit torque mechanism. GIXRD analysis confirms that increased Bi doping in Pt leads to progressive reduction in crystallinity, significant peak broadening, and preferred crystallographic orientation, indicating structural modifications in the Pt$_{100-x}$Bi$_{x}$ alloys. These findings are corroborated by cross-sectional TEM, which reveals a uniform distribution of Bi throughout the Pt layer and a concurrent loss of crystallinity of Pt. The non-linear dependence of $\theta_{\text{SH}}$ on $\sigma_{\mathrm{xx}}$ and $\rho_{\text{xx}}$ further reveals a $\rho_{\text{imp}}^{\text{SH}} \propto \rho_{\text{imp}}^2$ scaling, confirming extrinsic side-jump scattering as the dominant mechanism behind the enhanced SHE. Collectively, our findings establish a framework for energy-efficient spintronics beyond conventional $\mathrm{5}d$ transition metals. By demonstrating Bi alloying with Pt as an effective strategy for enhancing charge-to-spin conversion, our work offers design principles for next-generation neuromorphic computing, SOT-MRAM, Ising machines, and GHz spintronic devices.

\section{Experimental Section}

\subsection{Thin film preparation} High-resistance silicon (HR-Si) wafers ($> 10,000~\Omega\cdot\text{cm}$) were used as substrates. The Pt$_{100-x}$Bi$_{x}$(4.0 nm)/Co$_{40}$Fe$_{40}$B$_{20}$(3.0 nm) were grown on HR-Si/Ta(2.4 nm), and further protected with a capping layer of MgO(2.0 nm)/SiO$_2$(3.0 nm). Ta and Co$_{40}$Fe$_{40}$B$_{20}$ were sputtered at 11 W with 0.25 sccm Ar flow at a base pressure of \(4 \times 10^{-8}\) mbar, rising to \(4 \times 10^{-3}\) mbar during deposition. Other layers were deposited via electron beam evaporation. Pt$_{100-x}$Bi$_{x}$ was synthesized in situ by co-evaporation, with deposition rates monitored using a quartz crystal microbalance (QCM). The Bi source was stabilized before Pt adjustment. The deposition rate ratio required to achieve the desired Pt$_{100-x}$Bi$_{x}$ composition was calculated as:  \( \frac{v_a}{v_b} = \frac{M_a \rho_b}{M_b \rho_a} \cdot r \), where \( v_a \) and \( v_b \) are the deposition rates, \( M_a \) and \( M_b \) are the molar masses, \( \rho_a \) and \( \rho_b \) are the material densities, and \( r \) is the target molar ratio \cite{winkel2024comparative}.  \\

\subsection{Grazing Incidence X-ray Diffraction (GIXRD)} GIXRD measurements were used to determine the full-width-at-half-maximum (FWHM) of the diffraction peaks. Data were acquired using a Mat:Nordic SAXSLAB instrument with a Rigaku 003 microfocusing Cu X-ray source (parallel beam from a two-bounce monochromator) and two Dectris detectors: Pilatus3 300K R (orthogonal) and 100K (goniometer circle). The beam path was evacuated to suppress air scattering, and a $1^\circ$ incidence angle was employed to maximized signal intensity. Two-dimensional diffraction images were processed using SAXSGUI. FWHM values and lattice parameters were determined using TOPAS v6 (Bruker AXS, 2016), applying a pure platinum (Fm$\bar{3}$m) structural model and a pseudo-Voigt (PVII) profile. Peak broadening was analyzed separately. Exposure time was 3h per sample. Instrumental broadening was assessed using LaB6 powder.\\

\subsection{Cross-sectional high-resolution TEM and HAADF-STEM imaging}TEM measurements were carried out using a JEOL monochromatic ARM200F microscope, which is equipped with a Schottky field emission gun, a double-Wien monochromator, a probe aberration corrector, an image aberration corrector, a Gatan image filter (GIF) Continuum,  a Gatan Oneview camera, as well as a double silicon drift detector (SDD) for energy dispersive X-ray spectroscopy (EDS). The microscope was operated at 200 kV for the TEM measurements. TEM specimens were prepared by using a FEI Versa 3D focus ion beam–scanning electron microscope (FIB-SEM). After depositing a protection layer containing Pt and C using first the electron beam and then the Ga ion beam in the FIB-SEM, a lamella of the material was cut out using an ion beam at 30 kV and 1 nA. After transferring the lamella to a Cu TEM grid, the lamella was gradually thinned down by the ion beam. The thinning process was performed first at 30 kV and with a gradually decreasing beam current from 1nA to 100 pA. Then, the gentle polishing of the specimen was carried out with an ion beam energy of 5 kV and 2 kV to minimize ion beam effect. 

\subsection{Device Fabrication}ST-FMR bars (14$\times$4 $\mu m^2$) and SHNOs with width of 100 nm were fabricated via electron-beam lithography (Raith EBPG 5200), followed by Ar-ion milling. The GSG-CPW contact pads were subsequently fabricated by mask-less ultraviolet lithography (Heidelberg Instruments MLA 150) and a lift-off technique, followed by Cu(800 nm)/Pt(20 nm) electrodes, deposited by DC magnetron sputtering. Detailed fabrication process can be found in Ref.~\cite{kumar2022fabrication}.\\

\subsection{ST-FMR and Auto-oscillation (AO) measurements}For ST-FMR, signal generator (RSSMB 100 A)  was used to generate $I_{\text{rf}}$ and the ST-FMR voltage $V_{\text{dc}}$  was detected using a lock-in amplifier (SR830). For AO, $I_{\text{dc}}$ was applied using Keithley 2400 sourcemeter and AO signal was then amplified by a low noise Amplifier (LNF LNR4 14B), and observed using a spectrum analyser (RSFSV) with a resolution bandwidth of 1 MHz. All measurements were performed at room temperature.\\

\subsection{Statistical Analysis}The mean deviation (MD) quantifies dispersion in a dataset \( x_1, x_2, \dots, x_n \) and is computed with respect to the mean (\(\bar{x}\)) or median (\(M\)). The mean deviation about the mean is: $ MD_{\bar{x}} = \frac{1}{n} \sum_{i=1}^{n} |x_i - \bar{x}|$, where $\bar{x}$ is the arithmetic mean, $\bar{x} = \frac{1}{n} \sum_{i=1}^{n} x_i$. Here,  $|x_i - \bar{x}|$ represents the absolute deviation from the mean. \\

\medskip
\textbf{Supporting Information} 
Supporting Information is available from the corresponding authors upon request.

\medskip
\textbf{Acknowledgements} 
Financial support from the Horizon 2020 research and innovation programme (Grant Nos. 899559 “SpinAge” and 835068 “TOPSPIN”), and the Swedish Research Council (VR Grant No. 2024-01943), is gratefully acknowledged. Financial support from the Swedish Research Council (VR) and the Swedish Foundation for Strategic Research (SSF) for access to ARTEMI, the Swedish National Infrastructure in Advanced Electron Microscopy (Grant Nos. 2021-00171 and RIF21-0026), is also acknowledged. This work was performed in part at the Chalmers Material Analysis Laboratory (CMAL). We thank Dr. Michał Strach (CMAL) for his assistance and insightful discussions regarding the GIXRD measurements.

\medskip
\textbf{Conﬂict of Interest} \par
The authors declare no competing interests.

\medskip
\textbf{Authors' contributions} \par
 US, AK, TSP, MM and J{\AA} initiated the idea and planned the study. HH prepared the thin films with input from TSP and AK. AK fabricated the devices. US performed all electrical auto-oscillations, spin-orbit torque measurements, and analysis with input from AK and MR. LJZ and ABY performed TEM measurements with input from EO. J{\AA} and MM coordinated and supervised the project. All authors contributed to the data analysis and interpretation of the results and co-wrote the manuscript.

\medskip
\textbf{Data Availability Statement} \par
The data that support the findings of this study are available from the corresponding authors upon reasonable request.
\medskip

\bibliography{Main}

\end{document}